\documentclass[useAMS,usenatbib]{mn2e}
\pdfoutput=1
\usepackage{graphicx}
\usepackage{subfigure}
\usepackage{amsmath}
\usepackage{verbatim}
\usepackage{upgreek}
\usepackage{subfigure}
\usepackage[T1]{fontenc}
\usepackage{ae,aecompl}

\usepackage{color}

\title[A new HCN maser in IRAS\,15082$-$4808]{A new HCN maser in IRAS\,15082$-$4808}
\author[C. L. Smith, A. A. Zijlstra, G. A. Fuller]{C. L. Smith $^{1}$, A. A. Zijlstra $^{1}$, G. A. Fuller $^{1}$ \\
$^{1}$ Jodrell Bank Centre for Astrophysics, University of Manchester, Manchester, M13 9PL, UK}

\begin{document}

\date{Submitted XXXXXX Accepted XXXXXX}
\pagerange{\pageref{firstpage}--\pageref{lastpage}} \pubyear{2014}

\maketitle

\label{firstpage}
\begin{abstract}
We have identified a new vibrational HCN maser at 89.087 GHz in the asymptotic giant branch (AGB) star IRAS\,15082$-$4808, a maser which is thought to trace the innermost region of an AGB envelope. The observations of this maser at three epochs are presented: two positive detections and one null detection. The line profile has varied between the positive detections, as has the intensity of the maser. The major component of the maser is found to be offset by $-2.0\pm{0.9}$ km/s with respect to the systemic velocity of the envelope, as derived from the 88.631 GHz transition of HCN. Similar blueshifts are measured in the other 9 sources where this maser has been detected. Maser variability with pulsation phase has been investigated for the first time using the 10 stars now available. Comparisons with AGB model atmospheres constrain the position of the formation region of the maser to the region between the pulsation shocks and the onset of dust acceleration, between 2 and 4 stellar radii.

\end{abstract}


\section{Introduction}

Masers are found in a wide variety of astrophysical environments, both Galactic and extra-Galactic, and are a unique tool in the investigation of the dynamics, composition and structures found within these regions. OH and H$_2$O masers, for example, are found in active galactic nuclei (AGN) and luminous infra-red galaxies tracing accretion disks, radio jets and winds of an AGN (e.g. \citealp{Lo2005} and references therein). Within star formation regions, they can be used to probe outflows and disks as well as shocked regions of gas \citep{Caswell1995, Fish2007}.

Masers can also be used to study the circumstellar envelopes of evolved stars and planetary nebulae, with the chemistry and composition of the source dictating the observable species of maser. For instance, there are three types of maser often found in the circumstellar envelopes of oxygen-rich asymptotic giant branch (AGB) stars, namely SiO, H$_2$O and OH molecular transitions. In contrast, only one species of maser is commonly found in carbon-rich AGB stars. CO has been found to mase rarely, despite its high abundance in circumstellar envelopes. HCN, on the other hand, has been observed to mase in many different transitions \citep{Guilloteau1987, Schilke2003, Shinnaga2009}.

These masers allow us to probe very specific regions of the circumstellar envelopes of the AGB stars in which they reside. SiO masers trace regions close to the star that are under the influence of stellar pulsational shocks and magnetic fields \citep{Assaf2013, Perez-Sanchez2013}, whereas H$_2$O masers are found outside the dust formation zone and OH masers are found further still from the host star \citep{Cotton2008}. The 89.087 GHz HCN maser is thought to be formed in a small region close to the inner boundary of the circumstellar envelope of the host star, similar to the SiO masing region, due to the J=1-0 F=2-1 v=2 level lying 2050 K above the ground state and the maser spectral profiles' blue-shifted nature \citep{Goldsmith1988, Perez-Sanchez2013}. The HCN (J=1-0 F=2-1 v=2) 89.087 GHz maser therefore traces the region of the envelope which is currently undergoing acceleration. Having such a tracer is a useful tool in assisting in the understanding of envelope dynamics and wind acceleration.

To date, nine  HCN 89.087 GHz  masers have been detected in carbon-rich AGB stars \citep{Lucas1986, Guilloteau1987, Lucas1988, Lucas1989}. A number of these have been shown to vary on month time-scales, with some masers being completely undetected in subsequent observations and others having a change in line profile or intensity \citep{Lucas1988}. 

In this paper, we present a new detection of the 89.087 GHz HCN maser in IRAS\,15082$-$4808, along with an investigation of variation of maser intensity with stellar pulsational period.

IRAS\,15082$-$4808, also known as AFGL 4211, is a mass losing, carbon-rich AGB star. This source has been part of a number of line surveys (e.g. \citealp{Nyman1995, Woods2003}). Its mass loss rate has been derived from models to be approximately $1\times10^{-5}$ M$_{\sun}$ \citep{Groenewegen2002}. The distance to this source is not well constrained, but the adopted values vary between 640pc and 1500pc, although the general consensus is between 640 and 850pc \citep{Woods2003, Nyman1995}.

\section{Observations}

\subsection{Mopra 2010}\label{mopra2010}

The observations of IRAS\,15082$-$4808 were taken using the 22m ATNF Mopra telescope, 450km from Sydney, Australia, between the 18th June 2010 and the 27th June 2010 as part of a larger program.

The 3mm MMIC receiver (Monolithic Microwave Integrated Circuit) coupled with MOPS (the Mopra Spectrometer) in broadband mode were used, and covered the frequency range 84.5 to 92.5 GHz over four 2.2 GHz intermediate frequency bands (IFs). The IF of interest covered the range 88.5-90.5 GHz, putting this maser, at 89.087 GHz, more than 500 MHz from the edge of the band. The beam size at 90 GHz is 38" and the antenna efficiency is $\sim0.5$. The sensitivity of Mopra at 86 GHz is 22 Jy/K\footnote{{\emph{Technical summary of the Mopra radiotelescope}} (2005), www.narrabri.atnf.csiro.au/mopra/mopragu.pdf}.

The observations were reduced using the ATNF Spectral Analysis Package (ASAP). Observations that were badly affected by poor weather were identified and subsequently removed. The remaining sets of observations were averaged (weighted by the system temperature) and a polynomial baseline was fitted. The total integration time was 3 hours and the velocity resolution was 0.91 km/s. The rms noise of the resulting spectrum is 0.015 K in units of corrected antenna temperature and the $T_{\text{sys}}$ is 189 K. The spectra were inspected and lines identified using the Splatalogue and NIST databases \citep{NIST, Splatalogue}.

\subsection{Literature and archive data}

Following the identification, the literature and archives were searched for further sets of observations of IRAS\,15082$-$4808 in the required frequency range and two observations were found. The first were observations taken in August 1993 as part of the molecular line survey by \citet{Woods2003} using the Swedish-ESO Submillimeter Telescope (SEST, \citealp{Booth1989}) with rms noise level 0.014 K (main beam temperature units) and spectral resolution of 2.4 km/s. The sensitivity of SEST at 86 GHz is 25 Jy/K\footnote{{\emph{SEST parameters}} (1990), www.apex-telescope.org/sest /html/telescope-instruments/telescope/index.html}. The second are publicly available data from the Mopra archive and were observed in June 2011. The reduction of the SEST data is described in \citet{Woods2003}. The Mopra archive data was unpublished at the time of investigation, so the raw data files were taken from the Australian Telescope Online Archive and reduced in an identical manner to the Mopra 2010 data. The total integration time was 2.4 hours. The rms noise of the Mopra 2011 spectrum is 0.006 K in units of corrected antenna temperature, $T_{\text{sys}}$ is 157 K and the velocity resolution is 0.91 km/s.

\begin{figure}
\centering
\includegraphics[trim=2.5cm 13cm 2cm 2.5cm, clip=true,scale=0.48]{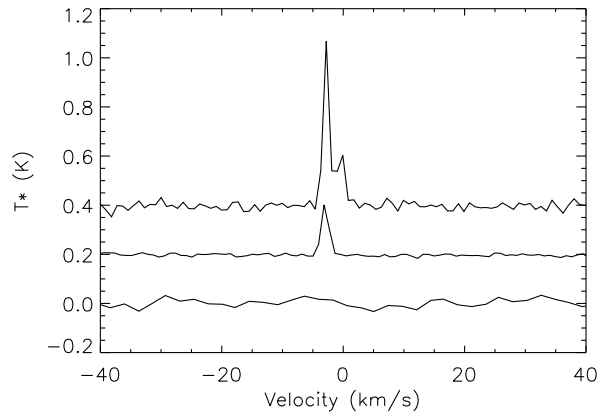}
\caption{HCN 89.087 GHz maser observations at three epochs. Velocity denotes the offset from LSR velocity in the frequency frame of the HCN maser. The upper spectrum shows the observation made in 2010 with Mopra, the middle spectrum shows the observation made in 2011 with Mopra and the lower spectrum shows the observation made in 1993 using SEST and is in units of main beam temperature. The Mopra observations are in units of corrected antenna temperature and the efficiency of Mopra at this frequency is $\sim$0.5. The sensitivities of Mopra and SEST at 86 GHz are 22 Jy/K$^1$ and 25 Jy/K$^2$ respectively. }\label{detection_plot}
\end{figure}

The maser was detected in the observations from 2011 and 2010 but was not detected in 1993. All resulting spectra are shown offset in Fig. \ref{detection_plot}. The line profile has varied significantly over the course of a single year, as is common for masers, going from an intense peak with a secondary component in 2010 to a significantly less intense peak with only a single detected component. \citet{Lucas1988} discuss the variability of this maser transition and indicate that it could be related to the pulsational period of the star as is the case with SiO masers \citep{Alcolea1999}.

\section{Analysis}

\subsection{Velocity}\label{analysis_velocity}

The maser in IRAS\,15082$-$4808 is non-gaussian in profile, with the Mopra 2010 observation having two distinct components. The HCN J=1-0 v=2 line is split into three hyperfine components: F=1-1 at 89.086 GHz, F=2-1 at 89.087 GHz and F=0-1 at 89.090 GHz. In order to identify whether the two components of the maser line are due to kinematic effects or multiple hyperfine components being observed, the velocities of the two peaks were compared with the theoretical separation of the hyperfine components of the J=1-0 v=2 transition. The velocities of the two observed components in the profile are taken at their respective peak values without fitting due to the line being unresolved. The major component and minor components are separated by 2.7$\pm$1.2 km/s which is significantly smaller than the separation between the hyperfine components of the HCN J=1-0 v=2 line which lie at separations of 5.02 km/s (between F=1-1 and F=2-1) and 7.46 km/s (between F=2-1 and F=0-1). Therefore, the double-peaked line profile is unlikely to be caused by a second masing hyperfine component, and it is more likely that it is caused by intrinsic motion of the masing molecules within the circumstellar envelope.

\begin{table*}
\centering
\caption{Known HCN 89.087 GHz maser sources with CO expansion velocities from \citet{Groenewegen2002} and maser velocity parameters taken from \citet{Lucas1988} and \citet{Guilloteau1987}. Mass-loss rates are taken from the reference listed in column 7. Velocity offset is defined as the difference between the velocity of the maser and the velocity of the shell with a negative value indicating a blue-shifted maser. In cases where the maser has multiple components, the velocity of the major component has been used.  The number of detections is the number of observations of the maser, excluding null detections. The period of pulsation is that derived from fitting the photometric data, as described in Section \ref{pulseperiod}.}\label{other_masers}
\begin{tabular}{l c c c c c c}
Source & V$_\text{exp}$ (km/s) & Velocity offset (km/s) & No. detections & Period (days) & $\dot{\text{M}}$ (M$_\odot$/yr) & Reference \\
\hline
CIT 6 & 17.5 & $-$5 & 7 & 637 & $6.5\times10^{-6}$ & 1\\   
IRC +10216 & 15.4 & $-$3 & 4 & 647 & $3.3\times10^{-5}$ &  1 \\
S Cep & 22.5 & $-$3.1 &  2 & 484 & $2.9\times10^{-6}$ & 1 \\
FX Ser & 28.4 & $-$1.1 & 2 & 531 & $3.1\times10^{-5}$ & 1 \\
IRC +50096 & 15.9 & $-$0.2 & 1 & 544 & $5.9\times10^{-6}$ & 1 \\
AFGL 2513 & 25.6 & $-$3.6 & 1 & 592 & $2.05\times10^{-5}$ & 2 \\
AFGL 2047$^\text{a}$ & 17.1 & $-$4.6 & 1 & 631 & $8.6\times10^{-6}$ & 3\\
IRC +30374 & 24.4 & $-$0.8 & 1 & 587 & $1.0\times10^{-5}$ & 4\\
T Dra & 16.8 & $-$0.7 & 1 & 424 & $8.2\times10^{-6}$& 1 \\
IRAS\,15082$-$4808 & 20.4 & $-$2.0 & 2 & 647 & $1.0\times10^{-5}$& 5\\
 \\
\end{tabular}

{\begin{flushleft}$^\text{a}$ Shell velocity taken as the mean of the central velocities of the CO(1-0) and HCN(1-0) lines from \citet{Nguyen1987}.

References: 1. \citet{Bergeat2005}, 2. \citet{Guandalini2006}, 3. \citet{Loup1993}, 4. \citet{Schoier2006}, 5. \citet{Woods2003}.
\end{flushleft}} 
\end{table*}

The HCN 89.087 GHz maser has been observed in nine other sources to date, which are listed in Table \ref{other_masers} along with this new source. The 89.087 GHz maser is blue-shifted by  up to 5 km/s with respect to the host star in all nine previously known sources \citep{Guilloteau1987, Lucas1988, Lucas1989}. 

To compare the velocity shift of the maser in IRAS\,15082$-$4808 to those observed in the other nine sources, we have taken the ground state, non-masing HCN 88.631 GHz (J=1-0, F=2-1) emission line, also observed as part of the larger program mentioned in section \ref{mopra2010}, as reflecting the systemic velocity of the envelope as was done by \citet{Lucas1988}. However, the line observed at this frequency at Mopra in the configuration described previously, is a blend of the hyperfine components of HCN: F=2-1, F=1-1 and F=0-1. These components must therefore be fitted in order to derive an accurate systemic velocity.

Using the methodology of \citet{Fuller1993}, we assume that the velocity profile is well-described by a gaussian distribution, and thus the total optical depth, $\tau$, of the J=1-0 transition may be expressed as:

\begin{equation}
\tau(v)=\tau\sum\limits_{i=1}^{3}\text{a}_i\exp{\left(-\frac{(v-v_0+v_i)^2}{2\sigma^2}\right)}
\end{equation}

\noindent where a$_i$ is the relative intensity of transition $i$ (a$_{F \text{upper}}$: a$_{2}$=1.0, a$_{1}$=0.6, a$_{0}$=0.2), $\sigma$ is the dispersion and $v_i$ is the offset velocity of transition $i$ with respect to a reference velocity, $v_0$, taken as the velocity of the central hyperfine component. The line profile can then be determined from:

\begin{equation}
T(v)=A(1-\text{e}^{-\tau(v)})+C
\end{equation}

\noindent where $A$ is the amplitude and $C$ is a constant to fit for any baseline offset that may exist.

\begin{figure}
\centering
\includegraphics[trim=2.5cm 13cm 2cm 2.5cm, clip=true,scale=0.48]{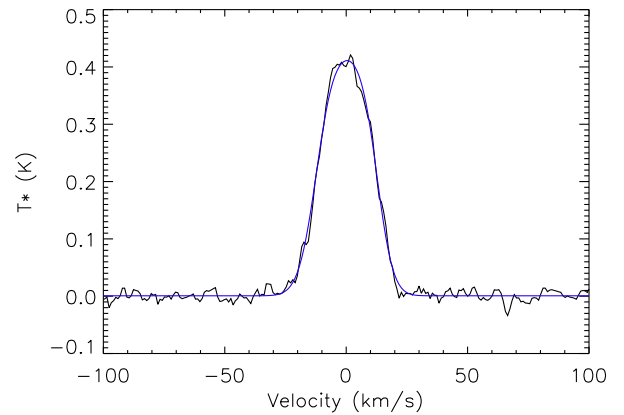}
\caption{HCN 88.631 GHz line as a blend of the F=2-1, F=1-1 and F=0-1 hyperfine components and the best-fit to the data is shown in blue. Velocity denotes the offset from LSR velocity in the frequency frame of the HCN 88.631 GHz line. T* is the corrected antenna temperature and the efficiency of Mopra at this frequency is $\sim$0.5. The sensitivities of Mopra and SEST at 86 GHz are 22 Jy/K$^1$ and 25 Jy/K$^2$ respectively. }\label{HCNnonmaser}
\end{figure}

The above model was fitted to the data using a hybrid method, based upon the use of the genetic algorithm {\sc{Pikaia}} \citep{Charbonneau1995} followed by the non-linear least squares fitting routine, {\sc{MPFit}} \citep{Markwardt2009}. The uncertainties on the fitted parameters were calculated by the latter from the covariance matrix. The result of the fitting is shown in Fig. \ref{HCNnonmaser} and gives the velocity of the central hyperfine component with respect to LSR velocity to be $-0.76\pm{0.06}$ km/s and the total optical depth of the transition as $1.28\pm{0.02}$. 

The velocity of the major observed component of the maser is offset with respect to the systemic velocity by $-2.0\pm{0.9}$ km/s (blue-shifted) and the minor component is offset with respect to the systemic velocity by $+0.7\pm{0.9}$ km/s (red-shifted).

\begin{figure*}
\centering
\subfigure{\includegraphics[trim=2cm 19cm 2cm 2cm, clip=true, scale=0.7]{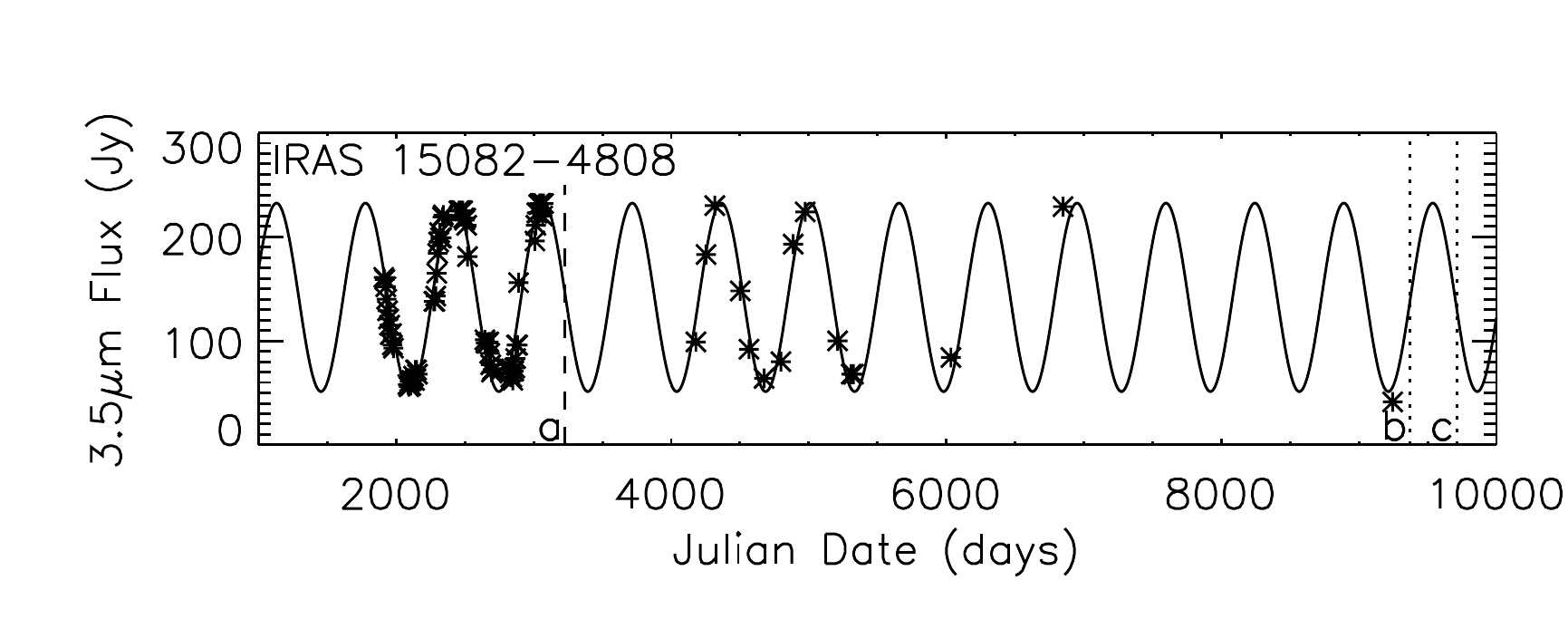}}
\subfigure{\includegraphics[trim=2cm 19cm 2cm 2.5cm, clip=true, scale=0.9]{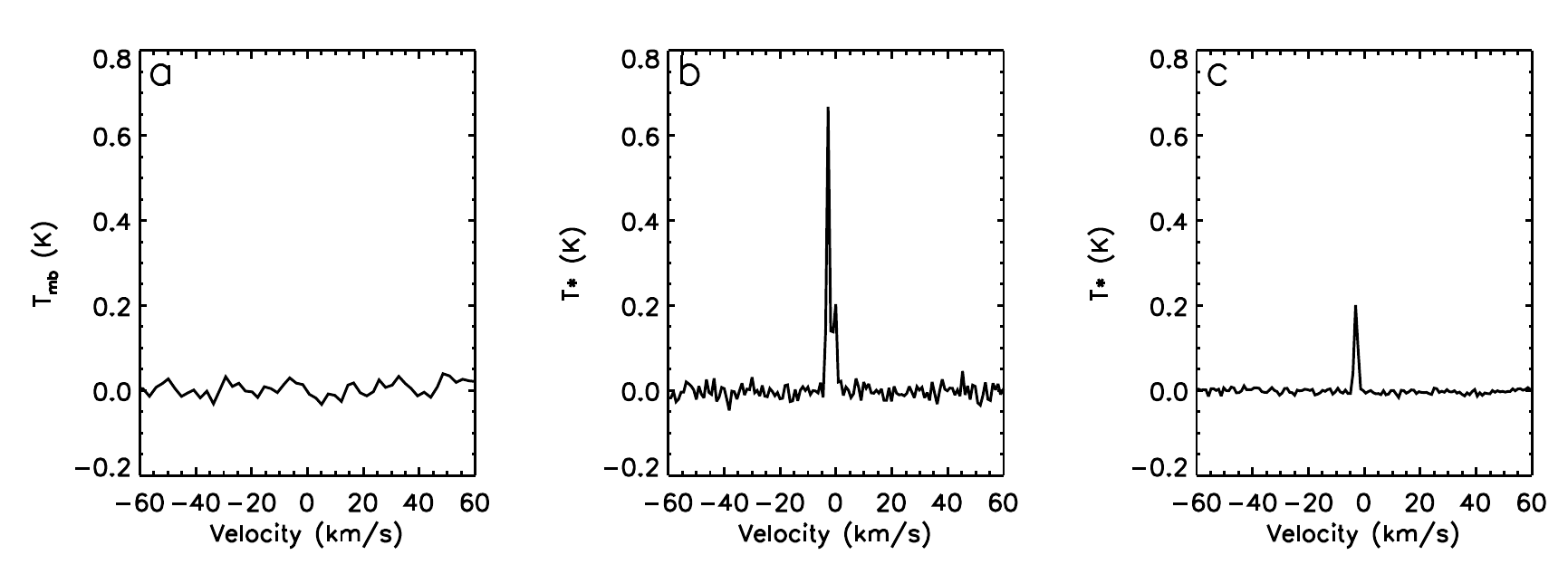}}
\caption{Pulsational cycle shown using 3.5 micron photometry. Photometric data taken from \citet{WISE}, \citet{Whitelock2006} and \citet{2mass}. The vertical dashed lines indicate the dates when the spectral observations were taken, and below are the spectra at each date. The spectrum labelled `a' is taken from \citet{Woods2003}.}\label{variability}
\end{figure*}

\subsection{Maser pumping}

\begin{figure}
\includegraphics[trim=2.5cm 13cm 2cm 2.5cm, clip=true,scale=0.48]{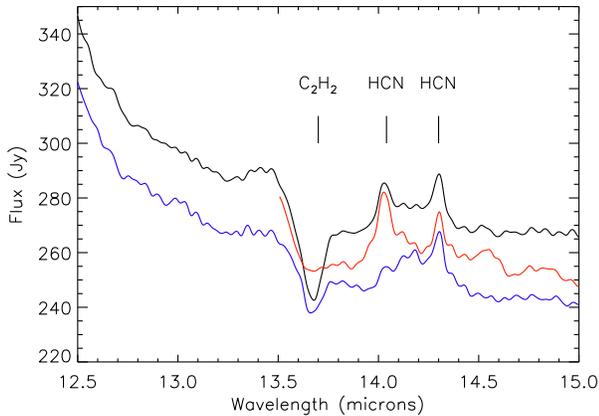}
\caption{SWS spectra from \citet{Sloan2003} showing the HCN and C$_2$H$_2$ features around 14 $\upmu$m. IRC+50096 is shown in black, T Dra (multiplied by a factor of 3) is shown in blue and the partial spectrum of S Cep is shown in red.} \label{SWS14}
\end{figure}

\begin{figure}
\includegraphics[trim=2.5cm 13cm 2cm 2.5cm, clip=true,scale=0.48]{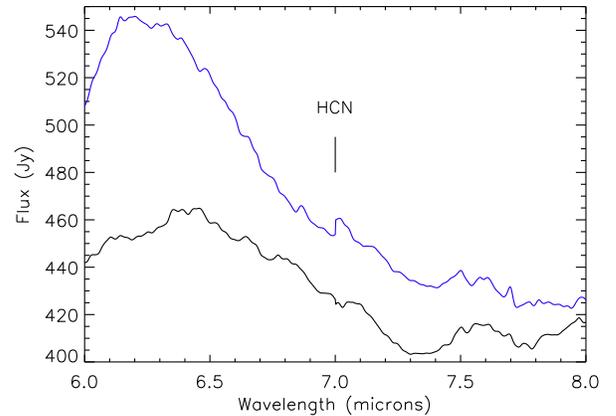}
\caption{SWS spectra from \citet{Sloan2003} showing the region of the 7 $\upmu$m feature. IRC+50096 is shown in black and T Dra (multiplied by a factor of 2.7) is shown in blue.}\label{SWS7}
\end{figure}

The velocity by which the maser line in IRAS\,15082$-$4808 is blue-shifted is significantly less than the expansion velocity of the envelope (20.4 km/s, {\citealp{Groenewegen2002})}, suggesting that the maser is being formed in a distinct region of gas below the wind acceleration zone. The red-shifted component of the maser in IRAS 15082$-$4808 has a velocity that is within uncertainties of the systemic velocity of the envelope. The lack of an observed definite red-shifted component suggests that either the maser is formed close enough to the star that the red-shifted component is obscured by the host star itself or that the maser amplifies photons emanating directly from the photosphere. These would both be in agreement with the interferometric observations of CIT 6 and IRC+10216 by \citet{Carlstrom1990} and \citet{Lucas1992} which suggest that the maser is located within 5 stellar radii.

As shown by \citet{Bujarrabal1981} and  \citet{Lockett1992}, SiO masers are formed close to the photosphere of the star but are amplified tangentially to the surface rather than radially, resulting in the maser emission being centred on the systemic velocity of the stellar envelope. In contrast, the blue-shifted central velocity of the 89.087 GHz maser with respect to the systemic velocity of the envelope suggests that the HCN maser is amplified radially, as is the case for OH masers, albeit significantly further from the stellar surface than is the case for this HCN maser. This difference in amplification direction between SiO masers and HCN masers means that the latter can be used to study the outflow velocities, whereas SiO masers cannot. 

\citet{Ziurys1986} show that it is possible for this maser to be pumped by the $\sim$14 $\upmu$m and/or the 7 $\upmu$m infra-red photons emitted by the star from the $v_2$ bands of HCN. The features at 14.0 $\upmu$m originate from the $2v_2^2-1v_2^1$ and $2v_2^0-1v_2^1$ transitions of HCN, the 14.3 $\upmu$m  feature from the $1v_2^1-0v_2^0$ transition and the 7 $\upmu$m feature from the $2v_2^0-0v_2^0$ transition (for further details see \citealp{Aoki1999}). {\emph{ISO}} SWS spectra from \citet{Sloan2003} were available for three of the previously known HCN maser sources from Table \ref{other_masers} (IRC+50096, T Dra and S Cep).  All three sources exhibit some degree of 14.0 $\upmu$m and 14.3 $\upmu$m emission  and their spectra in this region are shown in Fig. \ref{SWS14}. Emission in this band is suggested by \citet{Aoki1999} as being fairly common in carbon stars, and as shown by  \citet{Cernicharo1999} it is possible for the 14.3 $\upmu$m emission peak to be pumped radiatively from the ground state to the $v_2$ state.  Fig. \ref{SWS7} indicates that there was little-to-no 7 $\upmu$m emission in either T Dra or IRC+50096 when the ISO observations were made, which supports the 14 $\upmu$m pumping mechanism, and S Cep has no spectrum available at this wavelength.

The collisional rate coefficients of HCN $v_2=2$ are not yet published and therefore collisional excitation mechanisms cannot yet be ruled out. This should be addressed by modelling once these parameters are available.

\subsection{Relation to variability}\label{pulseperiod}

To determine the relation between variability and maser emission in IRAS\,15082$-$4808, the pulsational cycle was investigated using literature photometry \citep{WISE, COBE,Whitelock2006,  2mass}. 3.5 $\upmu$m photometry was collected and a sinusoidal curve fitted using the genetic algorithm {\sc{pikaia}} \citep{Charbonneau1995}. The resulting variability curve is shown in Fig. \ref{variability}.

\begin{figure*}
\begin{minipage}{\linewidth}
\centering
\includegraphics[trim=2cm 19cm 2cm 2.5cm, clip=true, scale=0.66]{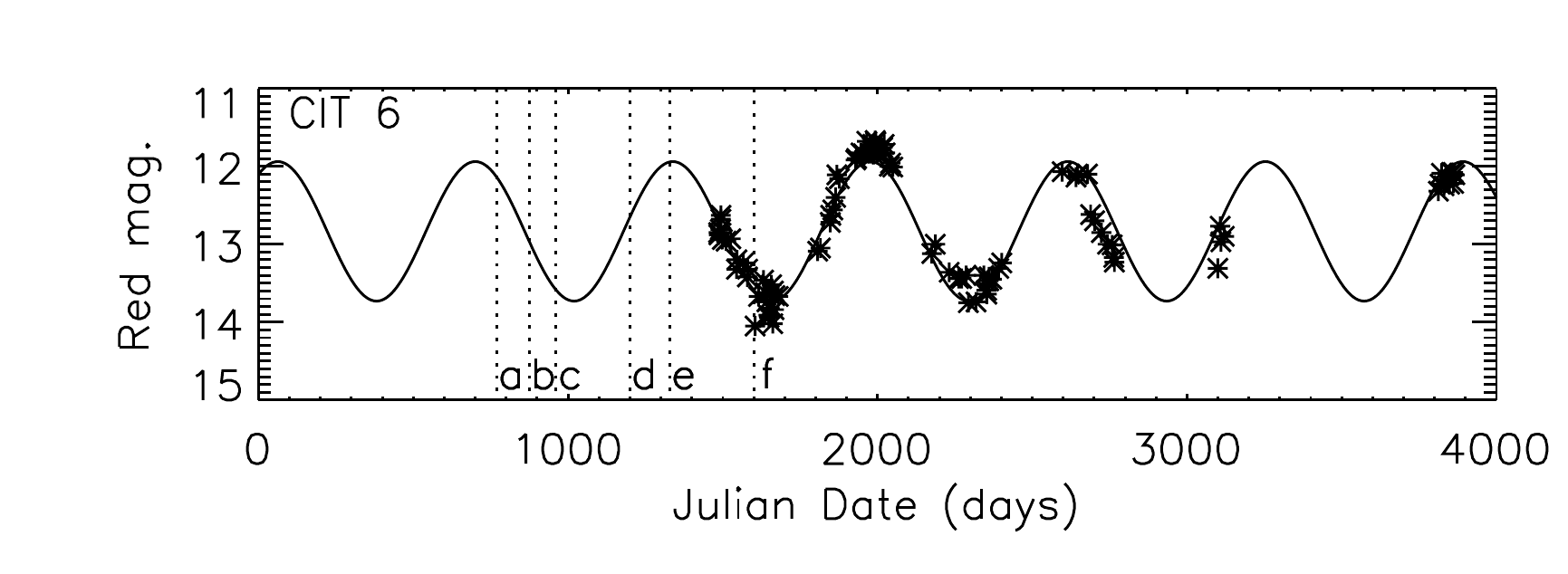}
\end{minipage}
\begin{minipage}{\linewidth}
\centering
\includegraphics[trim=2cm 19cm 2cm 2.5cm, clip=true, scale=0.66]{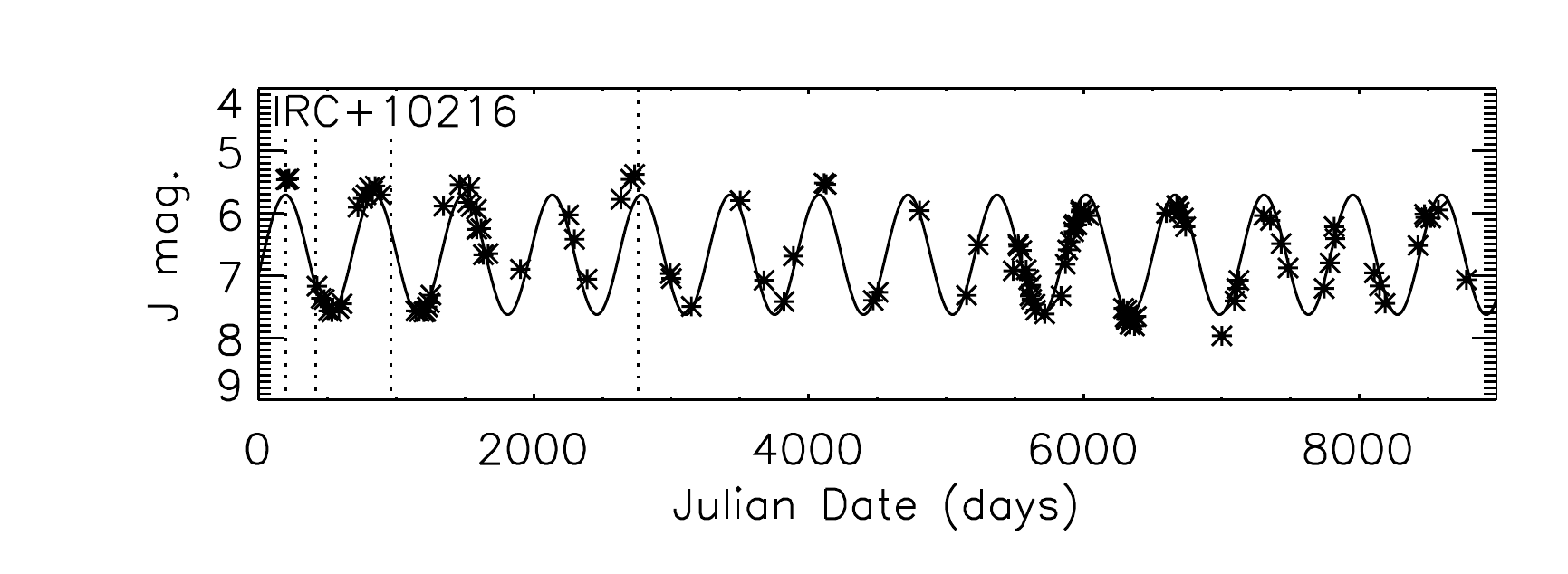}
\end{minipage}
\begin{minipage}{\linewidth}
\centering
\includegraphics[trim=2cm 19cm 2cm 2.5cm, clip=true, scale=0.66]{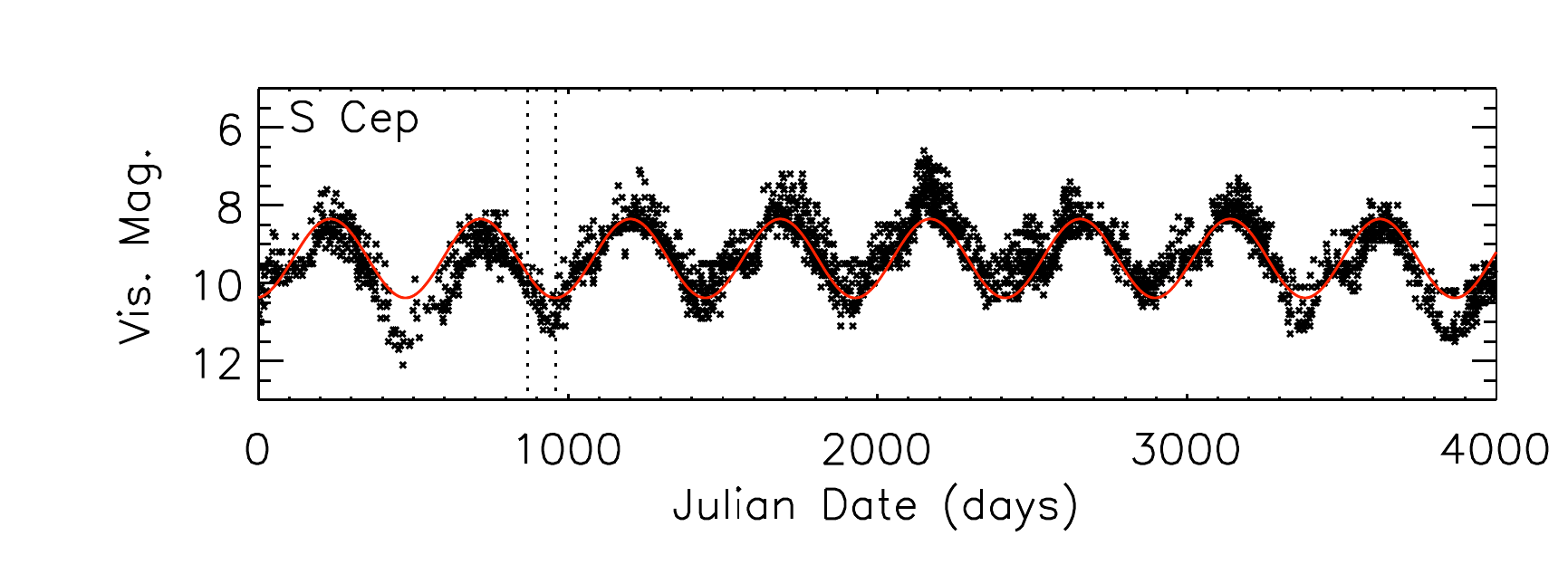}
\end{minipage}
\begin{minipage}{\linewidth}
\centering
\includegraphics[trim=2cm 19cm 2cm 2.5cm, clip=true, scale=0.66]{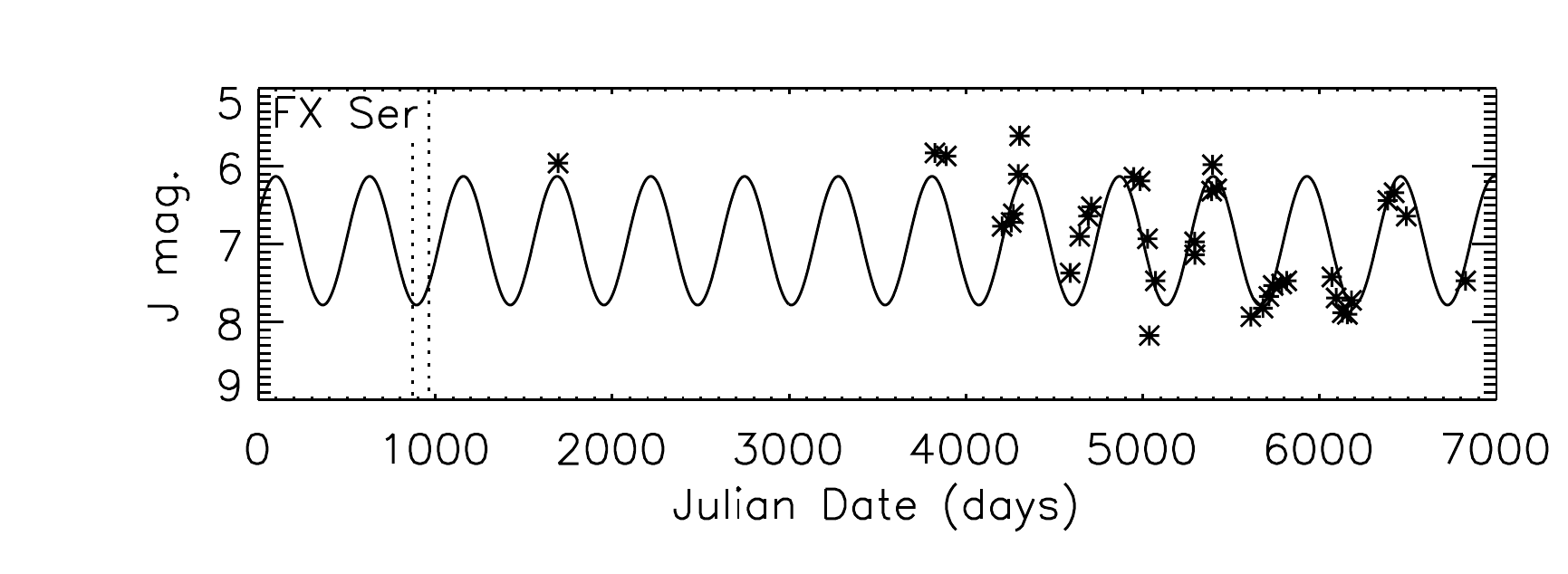}
\end{minipage}
\begin{minipage}{\linewidth}
\centering
\includegraphics[trim=2cm 19cm 2cm 2.5cm, clip=true, scale=0.66]{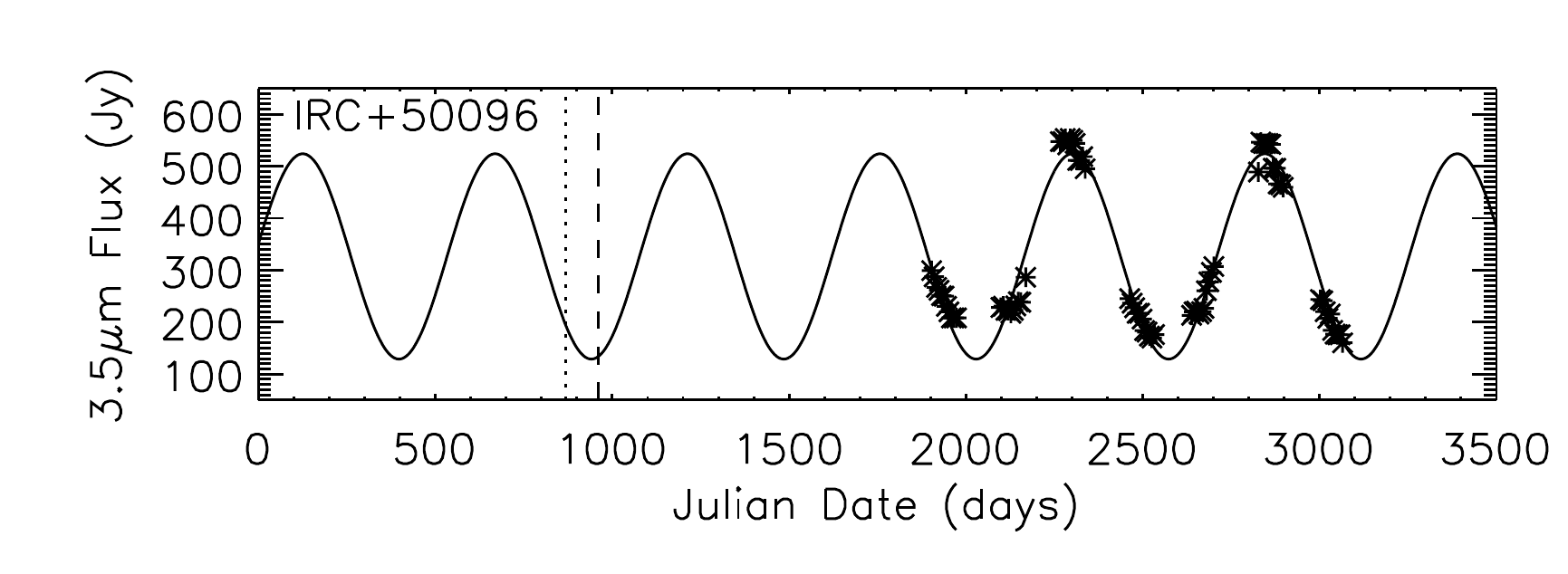}
\end{minipage}
\caption{Photometry data as a function of (Julian Date - 2446000) days. Photometry data taken from: \citet{Alksnis1995, Lebertre1992, 2mass, Whitelock2006, Shenavrin2011, AAVSO, Fouque1992,DENIS, Kerschbaum2006, COBE, IRAS,Jones1990,Epchtein1990}. Maser spectroscopic data taken from: \citet{Lucas1986, Guilloteau1987, Goldsmith1988, Lucas1988, Lis1989, Carlstrom1990, Lucas1992}. The three photometric data sets of IRC+10216 were offset from one-another in intensity; in order to fit a single sine curve to the data, offsets were applied to two of the data sets. This was not applied to any other source. Dotted lines indicate positive detections and dashed lines indicate null detections of the 89.087 GHz HCN maser. A further detection of the maser in CIT 6 was made by \citet{Nguyen1988} but the date of observation and the spectrum obtained were unavailable in the literature thus this has not been included on the figure.}\label{variability2}
\end{figure*}

\begin{figure*}
\begin{minipage}{\linewidth}
\centering
\includegraphics[trim=2cm 19cm 2cm 2.5cm, clip=true, scale=0.68]{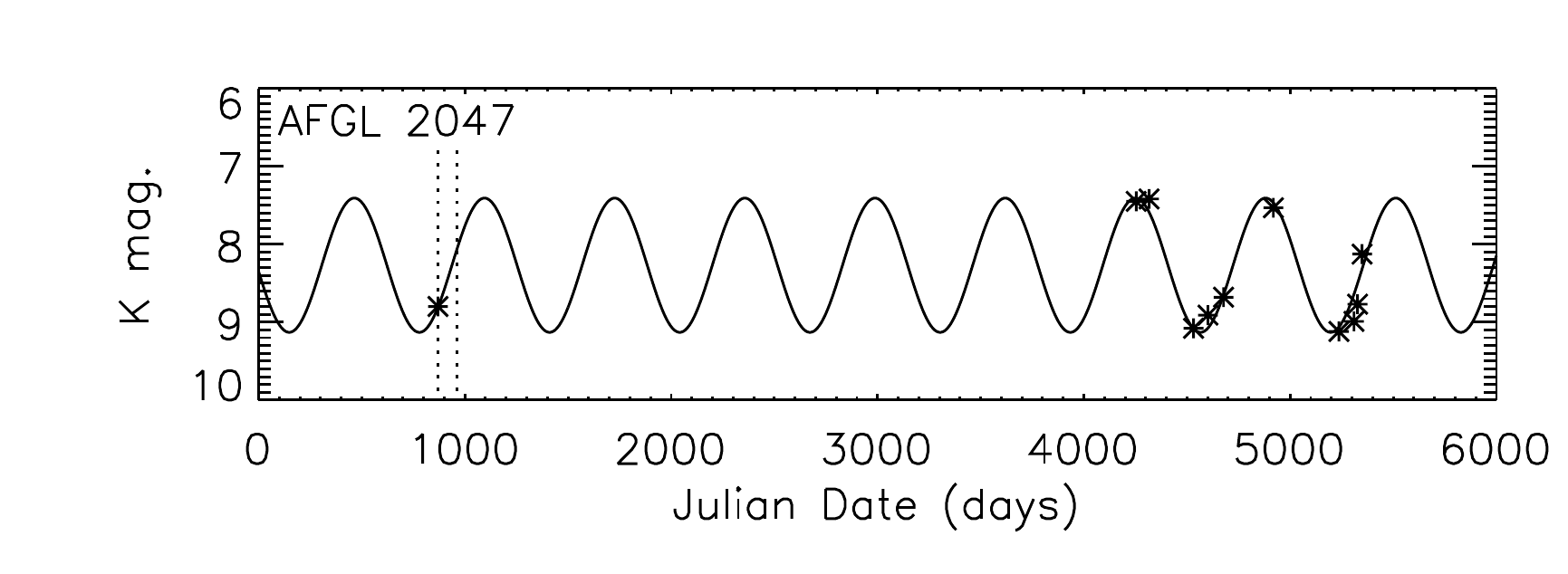}
\end{minipage}
\begin{minipage}{\linewidth}
\centering
\includegraphics[trim=2cm 19cm 2cm 2.5cm, clip=true, scale=0.68]{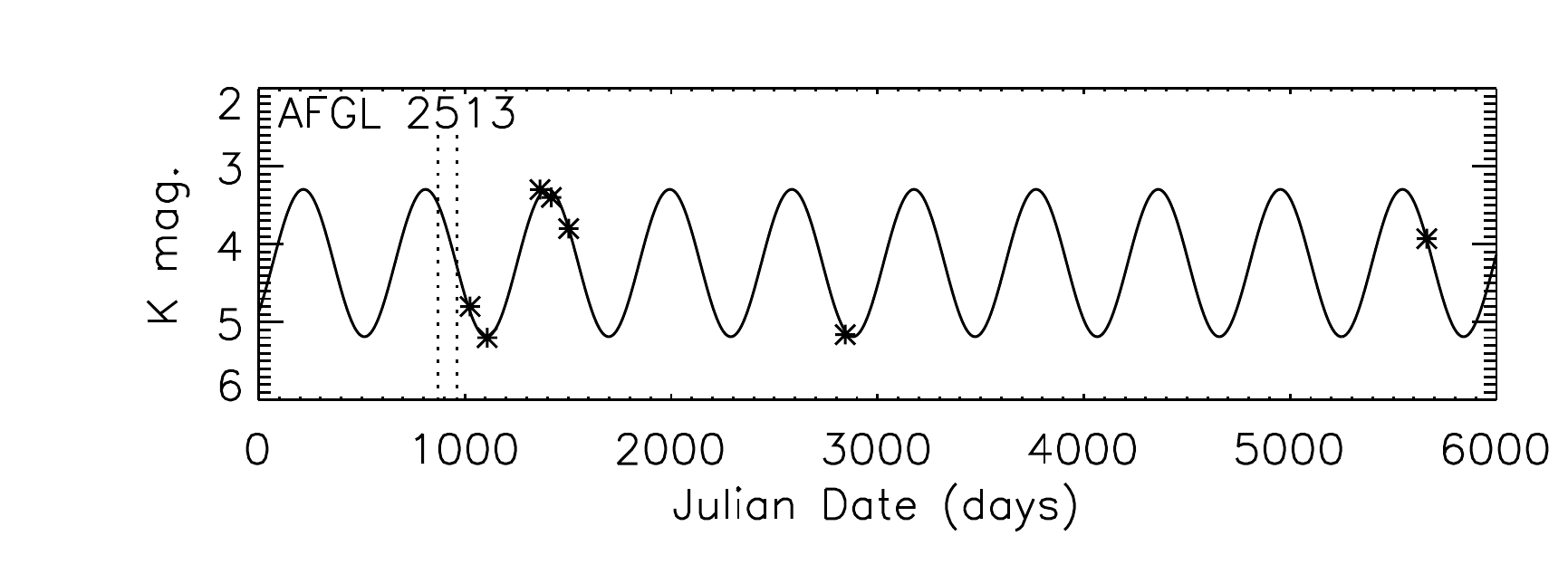}
\end{minipage}
\begin{minipage}{\linewidth}
\centering
\includegraphics[trim=2cm 19cm 2cm 2.5cm, clip=true, scale=0.68]{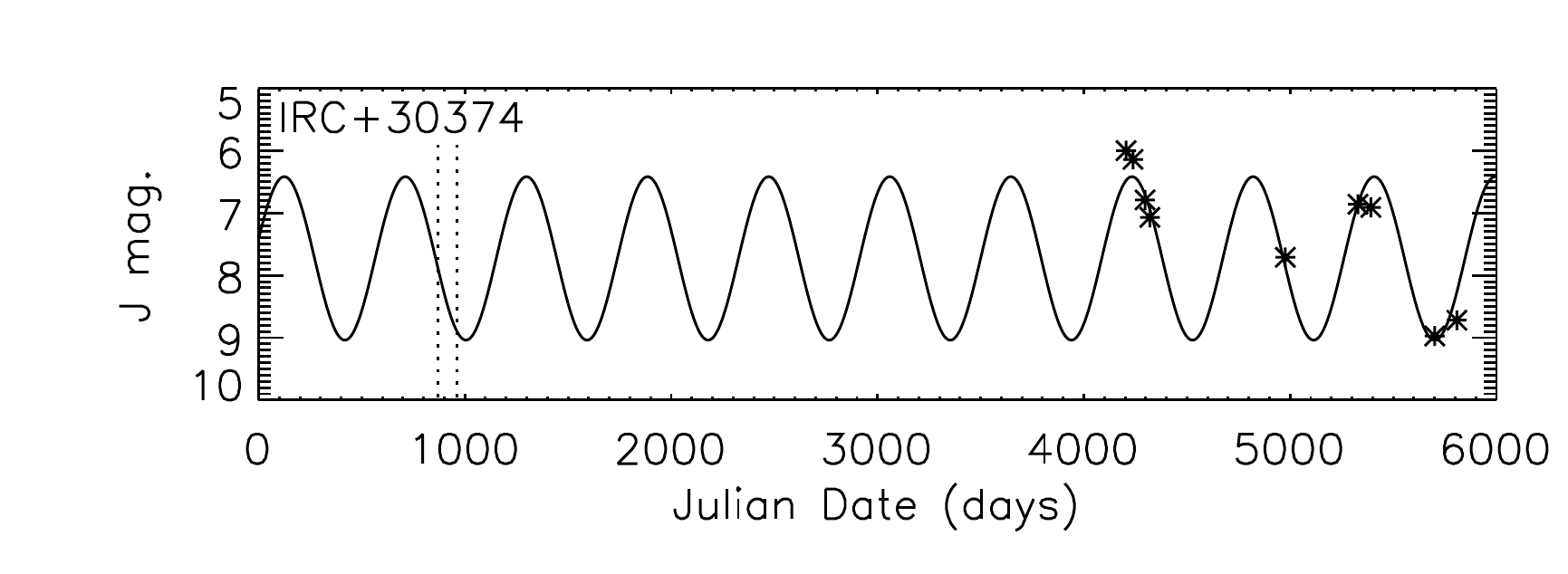}
\end{minipage}
\begin{minipage}{\linewidth}
\centering
\includegraphics[trim=2cm 19cm 2cm 2.5cm, clip=true, scale=0.68]{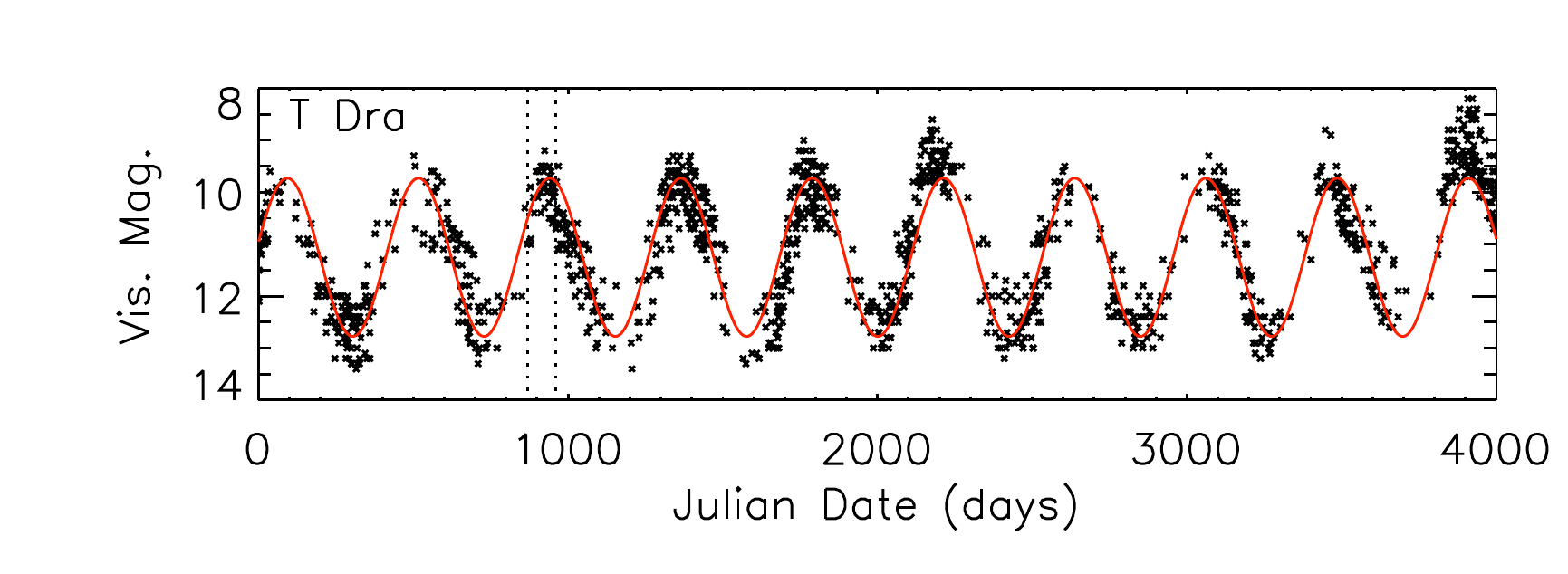}
\end{minipage}
\contcaption{}
\end{figure*}

The 1993 observation was clearly taken mid-way between a maximum and minimum brightness of the star. The 2011 observation was taken at a very similar point in the pulse phase, although there is little photometry taken around the time of this observation. However, if the pulsational period were the only cause of the maser variability and the pulsational cycle of IRAS\,15082$-$4808 has not changed significantly, then it would be expected that these two observations would have very similar intensities. This is not the case, with the 2011 observation clearly detecting the maser and the 1993 observation showing a null detection.

Variability curves were also created, following the same method as above, for each of the nine sources (Fig. \ref{variability2}). Dotted lines indicate positive detections and dashed lines indicate null detections. AFGL 2047, AFGL 2513, T Dra and IRC+30374 were observed either in March or June 1987 but it is not reported which and therefore both these dates are indicated on their respective variability curves.

The variability curves show that the HCN maser has been detected at a variety of phases in the stars' pulsational period, although some of the stellar curves have been extrapolated further than is reliable due to lack of photometry data. Most of the sources have only one published maser spectrum, despite multiple detections being reported in the literature, therefore it is impossible to determine whether the maser intensity or profile varies with period in these sources. 

\begin{figure}
\centering
\includegraphics[trim=3cm 13.2cm 0cm 2.8cm, clip=true, scale=0.5]{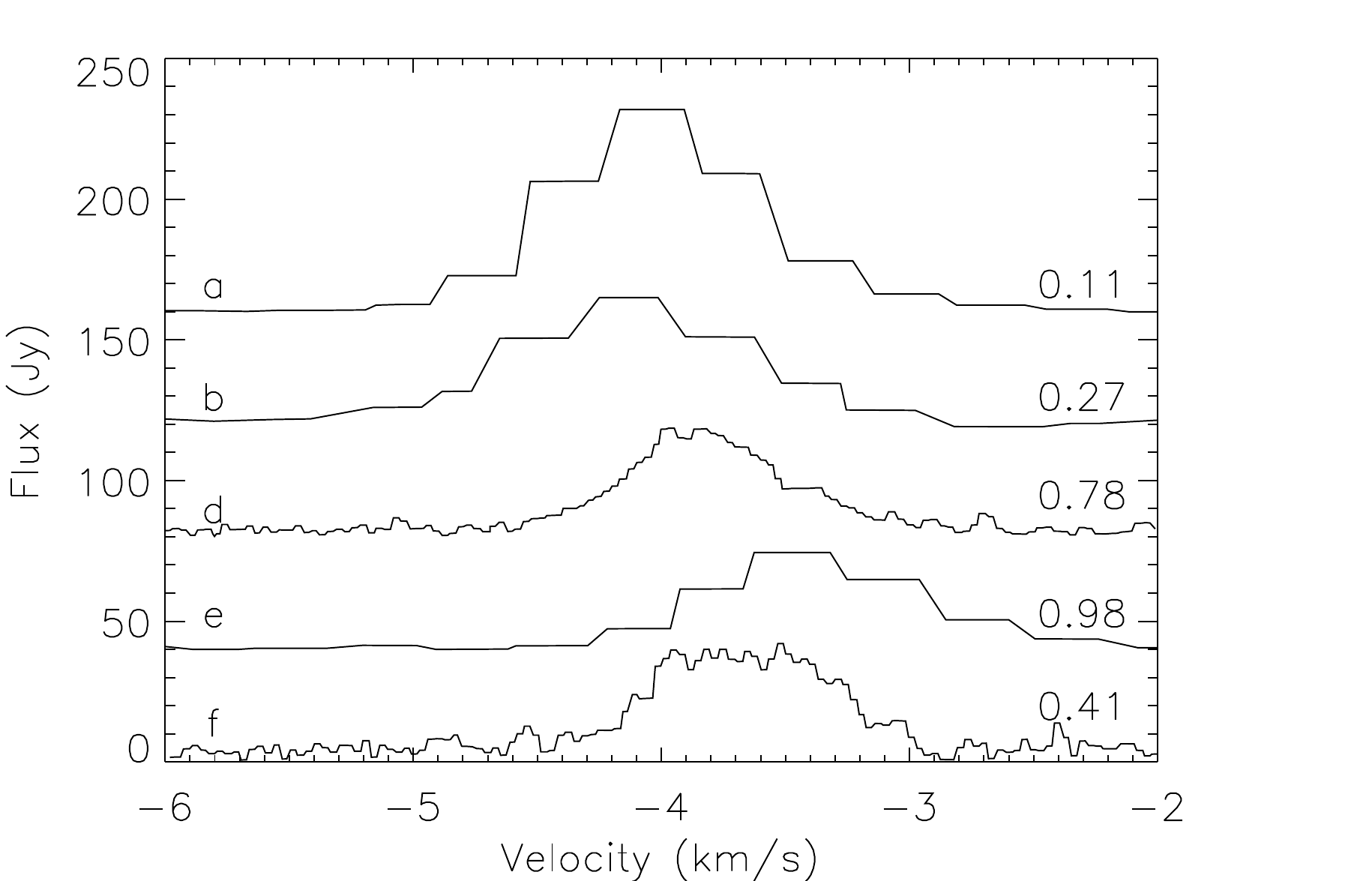}
\caption{CIT 6 spectra overlaid. \emph{a}, \emph{b}, \emph{d}, \emph{e} and \emph{f} correspond to the dates indicated on the upper plot of Fig. \ref{variability2}. Each spectrum is labelled on the right hand side of the figure with the stellar phase at the time of observation. A phase of 0.0 or 1.0 corresponds to stellar maxiumum and a phase of 0.5 corresponds to stellar minimum. Spectral data taken from \citet{Guilloteau1987, Goldsmith1988, Lucas1988, Lis1989, Carlstrom1990}. There is only a partial spectrum published for the observations made at date `c' and is therefore not included in this figure.}\label{CIT6}
\end{figure}

CIT 6, however, has seven  detections of the maser noted in the literature.  Five  of these have published spectra which are shown together in Fig. \ref{CIT6}.  There is a small change in central velocity and width of the line over the different observations, although these were made by a variety of telescopes at a variety of resolutions which could contribute to this. The intensity of the line has varied significantly over these observations: four  show similar peak intensities of approximately 40 Jy and the other approximately 75 Jy.

\begin{figure}
\centering
\includegraphics[trim=2.5cm 13cm 2cm 2.5cm, clip=true,scale=0.48]{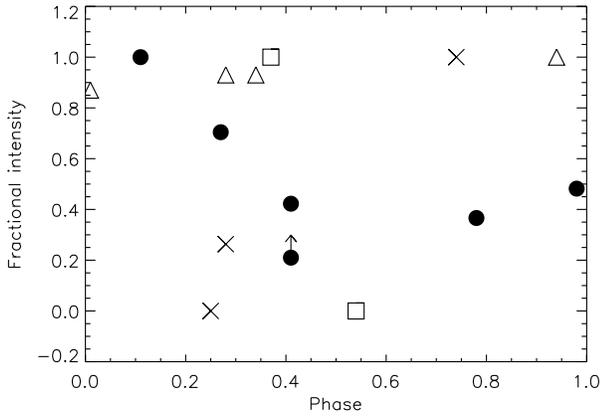}
\caption{Plot showing integrated intensity as a fraction of the strongest maser in each source against stellar pulse phase as calculated from the variability curves shown in Fig. \ref{variability} and Fig. \ref{variability2}. Filled circles: CIT 6, crosses: IRAS\,15082$-$4808, triangles: IRC+10216, squares: IRC+50096.}\label{phase_variability}
\end{figure}

To examine this variation with phase more closely, the integrated intensities of the CIT 6 lines were measured and plotted against phase, as derived from the sinusoidal fitting. The results are shown in Fig. \ref{phase_variability}, along with all other sources with multiple detections. The intensities were normalised to the maximum observed intensity of the maser in each source to allow comparisons between sources to be made. Where null detections were made, an intensity of zero is plotted. A phase of 0.0 or 1.0 indicates maximum intensity and a phase of 0.5 indicates minimum intensity.

CIT 6 is shown to be most intense close to stellar maximum and at lower intensities close to stellar minimum. This is somewhat in agreement with  IRC+50096 whose maser was detected at a phase of 0.37 and then not detected at a phase of 0.54. IRC+10216 was observed to have similar intensities and profiles at all epochs of observation \citep{Lucas1986, Guilloteau1987, Lucas1988, Lucas1992}. IRAS\,15082$-$4808 as discussed above, shows the intensity increasing between the phases of 0.2-0.7.

The 89.087 GHz maser in CIT 6 has been shown by \citet{Goldsmith1988} to exhibit linear polarisation of $20\%$. The observations were taken by a variety of instruments at a variety of elevations so this polarisation will have an effect on the observed intensity of the maser. For all other sources, polarisation has not been measured. The cause of this polarisation is currently unknown and further measurements are required in order to ascertain the cause \citep{Goldsmith1988, Perez-Sanchez2013}. Due to this, it is difficult to draw firm conclusions about variability with pulsation phase. In order to examine this more closely, a long-term measurement of maser intensity and polarisation in collaboration with photometric monitoring would need to be carried out over at least one pulsational period. From such a study, the effect of stellar variability on the intensity of the maser, the correlation between the two and, if mapped, the position of this maser with respect to the host star could be firmly established. This has been carried out successfully for SiO masers in several AGB stars (e.g. \citealp{Gonidakis2010}), thus a comparison between the two species of maser could also be made.

\subsection{Applications to AGB winds}

\citet{Richards2012} have examined water masers in oxygen-rich AGB envelopes. As part of their investigation, they studied the expansion velocities of the maser shells with respect to distance from the host star: Fig. 44 of \citet{Richards2012}. 

The expansion velocities obtained for the inner water maser shells at less than 20 AU from the star are consistent with the central velocity of the HCN masers with respect to the systemic velocities of their envelopes. A distance from the star of less than 20 AU is also consistent with current thoughts about the formation distance of the HCN masers: \citet{Carlstrom1990} report an upper limit of 180 AU on the size of the masing region in CIT 6 and interferometric observations by \citet{Lucas1992} suggest that the maser in IRC+10216 is located within 5 stellar radii. These imply that the HCN masers are tracing the expansion velocity of the region of the envelope in which they reside. 

The region in which HCN $v_2=2$ masers are thought to be formed in is very distinct. Thus the expansion velocity of a specific region inside the acceleration zone is represented by the velocity of the maser. \citet{Cernicharo2011} report detections of more than 60 vibrational transitions of HCN in IRC+10216, including the masing 89.087 GHz transition. That work showed a correlation between the excitation energy of the upper level and the observed line widths exists and suggests that this is due to different transitions emanating from different regions of the circumstellar envelope. This places the HCN $v_2=2$ maser in IRC+10216 within the 2-5 R* region of the envelope, which is consistent with the works of \citet{Carlstrom1990} and \citet{Lucas1992}. 

\begin{figure}
\includegraphics[trim=2.5cm 13cm 2cm 2.5cm, clip=true,scale=0.48]{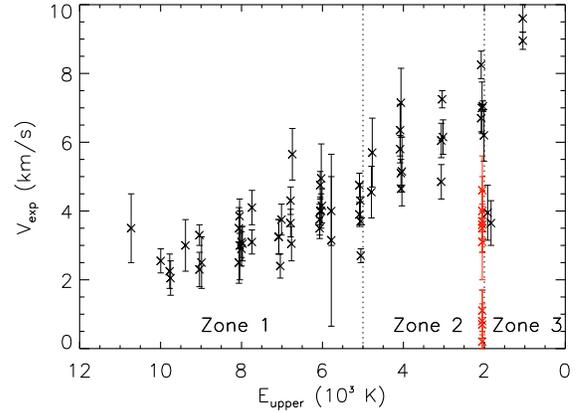}
\caption{Expansion velocity against the energy of the upper state (K) of a number of vibrational transitions of HCN. Expansion velocity is defined as $0.5\Delta\text{V}$ for the data taken from \citet{Cernicharo2011} (black points) and as the offset velocity with respect to the systemic velocity of the envelope for the ten HCN masers (red points). The zones are as calculated by \citet{Cernicharo2011}: zone 1 represents a region $\sim1.5$R*, zone 2 $\sim5$R* and zone 3 $\sim29$R*.} \label{EuVexp}
\end{figure}

Fig. \ref{EuVexp} shows the expansion velocities, taken as half of the measured linewidth, of the transitions reported in \citet{Cernicharo2011} against the energy of each upper state. Overplotted in red are the offset velocities of all ten detected HCN masers and these are found to be less than the velocities suggested by the velocity-E$_\text{upper}$ correlation. This suggests that the pumping mechanism for this maser may contain one of the high energy states found in zone 1. 

\subsection{Insights into AGB atmospheres}

Masers can provide insight into the winds of AGB stars through comparison of observed parameters with model extended atmospheres. The 89.087 GHz maser provides two significant constraints. First, there is very little velocity variability with pulsation phase, and no evidence for infall. This places the maser in a region of stable outflow. Second, the maser velocity is very low compared to the expansion velocity shown by CO. This locates the maser in a region where little acceleration has occured.

The dynamic model atmospheres  published in \citet{Hofner2003}, simulate time-dependent pulsations of the star for both mass-losing and non-mass-losing stars. Two models are discussed in detail in their Fig. 2 - 5, consisting of a star with parameters: 7000 L$_\odot$, 1.0 M$_\odot$, 355 R$_\odot$, a stellar temperature of 2800 K and a C/O of 1.4. One of these models is a mass-losing AGB star with mass loss rate $2.4\times10^{-6}$ M$_\odot$/yr, the other is non-mass-losing, and the pulsation period in both models, at 390 days, is significantly lower than that of any of the known maser sources. The parameters used for the two models differ only by the magnitude of the piston velocity amplitude and have values of 2 km/s and 4 km/s. A 2 km/s piston velocity amplitude is insufficient to cause mass-loss while 4 km/s allows dust production and subsequently mass-loss to occur. When these two models are compared with the parameters determined from the HCN 89.087 GHz masers, a number of insights can be made. 

\begin{figure}
\centering
\includegraphics[scale=0.9]{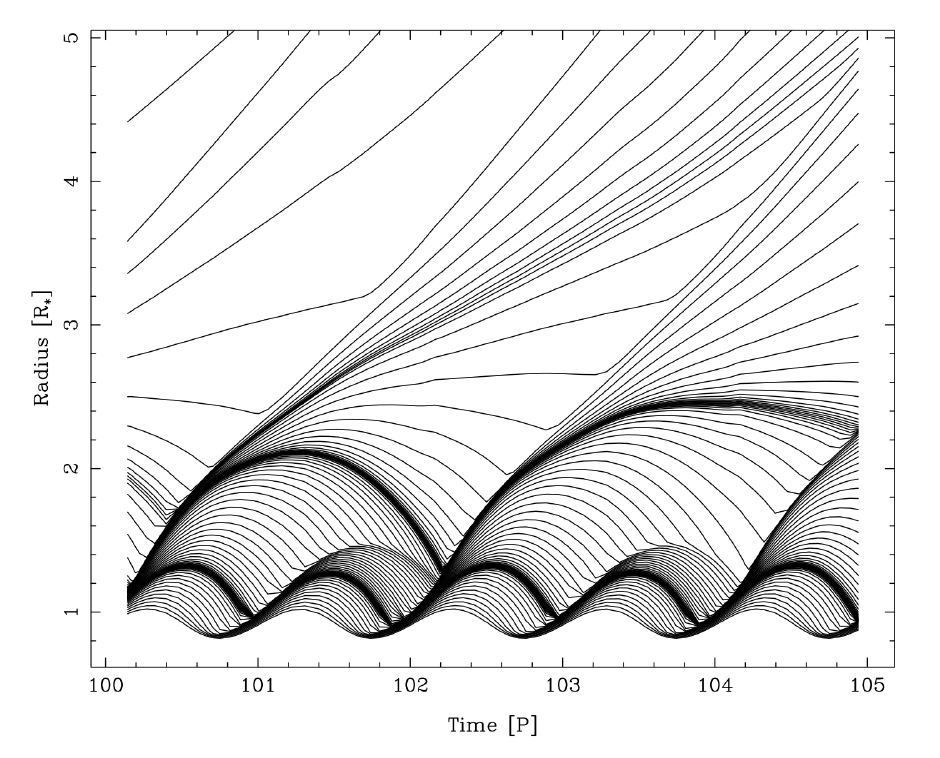}
\caption{Position of selected shells as a function of time for the mass-losing model 170t28c14u4, reproduced from Fig. 2 of \citet{Hofner2003},  A\&A, 399, 589 by kind permission of S. H{\"o}fner and ESO \copyright. } \label{hoefner_fig}
\end{figure}

From the aforementioned mass-losing model of \citet{Hofner2003}, it is clear that there are regions of the envelope undergoing regular expansion and contraction, clearly shown in Fig. \ref{hoefner_fig} (reproduced from \citet{Hofner2003} with kind permission from S H{\"o}fner and ESO \copyright ). Given the lack of correlation of maser intensity variability with pulse phase, it is unlikely that the maser would originate from one of these regions. The lack of an observed in-fall velocity suggests that the maser exists at the point that mass-loss begins. In model 170t28c14u4 of \citet{Hofner2003}, these requirements are satisfied at roughly 2 stellar radii from the host star, as can be seen in Fig. \ref{hoefner_fig}, coinciding with shocked regions of the envelope and in keeping with the expected distance of less than five stellar radii from \citet{Carlstrom1990} and \citet{Lucas1992}.

The lack of an observed substantial increase (more than a few km/s) in central velocity of the HCN maser lines indicates that this maser should not exist for long within a particular region of gas as it moves outwards. The implication is that the maser may cease operating as the region outwardly expands from the central star.

The upper level of the HCN 89.087 GHz maser lies at 2050 K above the ground state. The gas temperature in model 170t28c14u4 drops below this at roughly 1.5 stellar radii. The region of the model AGB atmosphere between 1 and 2 stellar radii has a highly variable flow velocity, changing with pulse phase ($-13$ km/s to $+10$ km/s). If the maser is radiatively pumped by photospheric photons, the maser will exist further from the host star than that dictated by the gas temperature. In the aforesaid model, at 2.5 to 3.5 stellar radii, the velocity is more stable with pulse phase  and, with flow velocities of 0 km/s to +8 km/s, more closely reflects the offset velocities of HCN masers. The gas temperature in this region is approximately 1000 K. 

The non-mass-losing model, 170t28c14u2, has flow velocities of between +1 and +4 km/s in the region of 1.1-1.2 stellar radii for two of four pulse phases and between $-11$ and $-13$ km/s for the remaining phases. In this region, the gas temperature is roughly 2000 K. At 1.2-1.4 stellar radii, the temperature drops to approximately 1000 K and the flow velocities range from -10 km/s to 10 km/s. The 89.087 HCN GHz masers are only observed in mass-losing carbon rich AGB stars, thus this model is not a suitable representation for these stars.

Neither of these models perfectly reproduce the conditions required for this maser to exist, but these models are not tailored to the individual sources and several model parameters vary significantly from those of the known host stars (e.g. pulse period). However, it is promising that the formation regions suggested from the mass-losing model agree  with the regions predicted in \citet{Carlstrom1990} and \citet{Lucas1992}.

\section{Conclusions}

We have identified a new 89.087 GHz HCN maser in the AGB star IRAS\,15082$-$4808. This maser was first detected in this source in June 2010 using the Mopra telescope in Australia, and was detected once again in June 2011 at Mopra. It was, however, undetected in a molecular line survey by \citet{Woods2003} based upon observations taken at SEST in 1993. 

The observation in 2010 has two components to the maser, and that of 2011 has only a single component at the same velocity shift as the major component in 2010. It is highly unlikely that the two component profile is due to a second hyperfine component being observed and is more likely that it is due to motions of the masing molecule within the stellar envelope. 

The major component of the maser in IRAS\,15082$-$4808 as observed in 2011 is offset by $-2.0\pm{0.9}$ km/s (blue-shifted) with respect to the systemic velocity as derived from the HCN 88.631 GHz emission line and the minor component by $+0.7\pm{0.9}$ km/s (red-shifted). The offset velocity of the major component is in-keeping with the $-0.3$ to $-5$ km/s offset of all previous detections of this maser in other sources \citep{Guilloteau1987, Lucas1988, Lucas1989}.

The variability of the maser in all ten sources in which it has been detected to date has been investigated, although firm conclusions could not be drawn due to lack of observations and polarisation effects. Long term monitoring of intensity and maser polarisation in conjunction with regular photometric measurements would allow the variation of maser intensity with stellar pulsation to be properly assessed.

Finally, comparisons of the derived maser parameters from all known host stars have been made to published models of AGB atmospheres. The comparison suggests that the masers are formed between 2 and 4 stellar radii from the host star and at the point that mass-loss begins. This region also contains shocked gas. Despite the gas temperature within this region being lower than that of the upper maser level, radiative pumping from photospheric photons would allow the existence of this maser within this region. The outflow velocities of the gas within this region are matched well by the offset velocities of the known masers. This region also agrees with the measurements and predictions of \citet{Carlstrom1990} and \citet{Lucas1992}, which place the maser within 5 stellar radii. High spatial resolution interferometry would be required to be able to identify for certain the exact region of the envelope in which this maser is formed.

\section{Acknowledgements}

We would like to thank J. Williams for taking the observations upon which this work is based and S. H{\"o}fner and ESO $\copyright$ for kindly allowing the reproduction of Fig. \ref{hoefner_fig}. We acknowledge with thanks the variable star observations from the AAVSO International Database contributed by observers worldwide and used in this research. The Mopra radio telescope is part of the Australia Telescope National Facility which is funded by the Commonwealth of Australia for operation as a National Facility managed by CSIRO. The University of New South Wales Digital Filter Bank used for the observations with the Mopra Telescope was provided with support from the Australian Research Council. This research was supported by the Science and Technologies Funding Council. 

\bibliography{Maser_refs}
\end{document}